# Non-ionizing cross section of electron scattering on atoms in matter accounting for dynamical screening effect

N. Medvedev[1,2,*], D.I. Zainutdinov[3], A.E. Volkov[3]

1. Institute of Physics, Czech Academy of Sciences, Na Slovance 1999/2, 182 21 Prague 8, Czechia
2. Institute of Plasma Physics, Czech Academy of Sciences, Za Slovankou 1782/3, 182 00 Prague 8, Czechia
3. P.N. Lebedev Physical Institute of the Russian Academy of Sciences, Leninskij pr., 53, 119991 Moscow, Russia

## Abstract

We present a model of non-ionizing scattering of electrons on atomic ensemble in matter, applicable in a wide electron energy range from ~eV up to relativistic ones. The approach based on the dynamic-structure factor formalism considers collective response of the atomic and electronic systems of a target. It accounts for dynamical screening of atomic nuclei in matter by valence (collective) and core-shell electrons during the scattering events, dependent on the incident electron velocity. The proposed formalism for the cross section enables us to describe in a unified manner the electron scattering on the ensemble of isolated atomic nuclei at high incident electron energies, reducing to the scattering on phonons with decrease of the energy. Our model can be used, e.g., in transport Monte Carlo codes to describe the energy exchange between excited electrons and atomic ensemble in matter. An example of swift heavy ion track formation in quartz simulated with the proposed cross section shows a reasonable agreement with the experiment validating the model.

[*] Email: nikita.medvedev@fzu.cz

# I. Introduction

Radiation transport in matter has a variety of applications from radiation safety, to space missions and medical treatment [1–4]. Electron irradiation is widely applied in electron diffraction and microscopy [5]. Synchrotrons and free-electron lasers use X- and γ-ray irradiation for studying structural and dynamical processes in matter and chemical reactions [6]. High-energy charged particles irradiation of materials is used in nanotechnologies and material modification [7,8]. Swift-heavy ion irradiation induces electron cascades along the ion trajectories followed by spatial spreading of the generated electrons, which leads to energy deposition to target atoms causing finally formation of observable and latent ion tracks [9,10].

In these irradiation scenarios, electron transport plays a crucial role [9,11]. At typical energies of incident particles, triggering an electron cascade, the electrons have energies from ~MeV down to ~eV energies and lower. A high energy electron experiences three main channels of interaction with matter realize [9,11–15]: (i) scattering on target electrons, exciting electrons from core atomic shells or valence/conduction band (impact ionization or excitation of collective electronic modes – plasmons); (ii) scattering on the atomic ensemble of the target without electron ionization (transferring energy to individual atoms or exciting their collective modes – phonons); (iii) photon emission *via* Bremsstrahlung process, which is important mainly at high relativistic energies.

Transport Monte Carlo (MC) simulations is the standard technic for modeling of electron transport and accompanying effects [9,12,16,17]. The scattering cross sections of projectiles are the key parameters of these simulations, describing the scattering probabilities of a particle, its mean free path, and energy loss in a target [9,11,14,15,18]. A number of works, see e.g. [19–24], studied the scattering of electrons on the electronic system of a target, accounting for collective response of the material.

Here, we will focus on the electron-atomic system scattering channel, in which the energy transfer only provides atoms of the solid with energy: non-ionizing scattering cross section on the atomic ensemble. Electron scattering on atoms in a condensed phase includes scattering on individual atoms and excitation of vibrational or rotational degrees of freedom in molecular case. In solids, there are two main mechanisms of electron-atom scattering: individual atom, and collective vibrational modes, phonons. Various conventions are used in different communities, in which sometimes both types of the scattering are called "elastic scattering", while in others the electron-phonon scattering is referred to as "quasi-elastic" or even "inelastic" scattering. In our work, we will call the both types of scattering with a unified term "e-a scattering".

Simulations of such scattering usually rely on approximations: low-energy electrons are assumed to scatter on collective atomic oscillations – phonons, whereas high-energy electrons interaction is assumed to be with individual and dynamically independent atoms [11,13–15,25]. The intermediate region, where neither of these approximations work (typically for electron energies in the range from a few eV up to a hundred eV), is not covered well with existing models, as well as amorphous or liquid materials which cannot be described with the phononic approximation. Taking into account the dependence of the screened charge of atoms on the electron velocity, we propose a model based on the dynamic-structure factor formalism, naturally unifying these limiting cases and covering a wide energy range of a scattering electron.

## II. Model

### 1. General form of the scattering cross section

To derive a general form of the scattering of an electron on a coupled system of nuclei and electrons forming a material of interest, we start with the first-order Born approximation, which assumes plane waves for an incident particle [26–28]. The double differential cross section of interaction of an electron is expressed as follows:

$$\frac{d^2\sigma}{d(\hbar q)d(\hbar\omega)} = \frac{q}{2\pi\hbar^3 v^2} \sum_{i,f} P_i |\langle f|\langle k_f^e|\hat{V}_{int}|k_i^e\rangle|i\rangle|^2 \delta(\hbar\omega - (E_f - E_i)). \tag{1}$$

Here $\hbar\omega$ is the transferred energy, $\hbar q$ is the transferred momentum (wave vector), with $\hbar$ being the Planck constant; $v$ is the incident electron velocity; $E_f$ and $E_i$ are, correspondingly, its final and initial energies; $|i\rangle$ and $|f\rangle$ are the initial and the final states of the target; $|k_{i,f}^e\rangle$ are the initial and final states of the incident electron, $\hat{V}_{int}$ is the interaction potential; the scattering matrix element is averaged over the initial target states with the factors $P_i$ (Gibbs or Boltzmann statistical weights of the medium states in thermal equilibrium) [26–28]. Here, an isotropic and uniform target medium is assumed.

The interaction of the electron with charged particles in the target (nuclei and electrons) is described by the pairwise Coulomb terms:

$$\hat{V}_{int} = -\sum_{j=1}^{N_i} \frac{Ze^2}{(\boldsymbol{r}-\boldsymbol{R}_j)} + \sum_{j=1}^{N_e} \frac{e^2}{(\boldsymbol{r}-\boldsymbol{r}_j)}, \tag{2}$$

here $R_j$ and $r_j$ are the coordinates of the nuclei and electrons, respectively; the nuclei have the charge $Ze$ ($e$ is the electron charge); the terms are summed over all nuclei $N_i$ and electrons $N_e$ of the target.

In the plane waves approximation for the incident and scattered electronic states ($|k_{i,f}^e\rangle = V^{-\frac{1}{2}}\exp(i\boldsymbol{k}_{i,f}^e\,\boldsymbol{r})$, with $V$ being the volume of the scattering system), the cross section then splits into the following sum [29]:

$$\frac{d^2\sigma}{d(\hbar q)d(\hbar\omega)} = \frac{q}{2\pi\hbar^4}\frac{1}{v^2}\left(\frac{4\pi e^2}{q^2}\right)^2 \left(S_{ee}(\omega,q) - ZS_{ei}(\omega,q) - ZS_{ie}(\omega,q) + Z^2 S_{ii}(\omega,q)\right), \tag{3}$$

where the partial dynamic structure factors are defined as [26–28]:

$$S_{ab}(q,\omega) = \sum_i P_i \left\langle i \left| \int \frac{dt}{2\pi} \exp(i\omega t) \int d\mathbf{r} \int d\mathbf{r}' \exp(-i\mathbf{q}(\mathbf{r}-\mathbf{r}')) \hat{n}_a(\mathbf{r},t)\hat{n}_b(\mathbf{r}',0) \right| i \right\rangle, \quad (4)$$

the indices {a,b} mark electrons (e) and ions (i), and the number-density operators are $\hat{n}_e(\mathbf{r},t) = \sum_{l=1}^{N_e} \delta(\mathbf{r} - \hat{\mathbf{r}}_l^e(t))$ and $\hat{n}_i(\mathbf{r},t) = \sum_{l=1}^{N_a} \delta(\mathbf{r} - \hat{\mathbf{R}}_l^i(t))$.

Following the idea of Chihara decomposition [30], we split the electronic contributions $S_{ie}(q,\omega)$, $S_{ei}(q,\omega)$ and $S_{ee}(q,\omega)$ into the core-shells and the 'free' electrons (belonging to the valence or the conduction band of the material). The core-shell electrons are assumed to follow their parenting ions adiabatically, whereas valence/conduction electrons are free particles influenced by the potential of ions. This allows dividing the general scattering cross section on matter in Eq.(3) into two parts: the first describes scattering on the coupled system of screened atomic nuclei, the second describes scattering on the correlated electron subsystem.

The *e-a* scattering cross section on the atomic ensemble (as defined above: scattering without ionization or electronic excitation of the target) can then be rewritten as follows [30]:

$$\frac{d^2\sigma_{e-at}}{d(\hbar q)d(\hbar\omega)} = \frac{q}{2\pi\hbar^4} \frac{1}{v^2} \left(\frac{4\pi e^2}{q^2}\right)^2 \left(Z - Z_I f_I(\tilde{q}) - \rho(\omega,\tilde{q})\right)^2 S_{ii}(\omega,q). \quad (5)$$

Here $Z - Z_I f_I(\tilde{q}) - \rho(\omega,\tilde{q})$ is the charge of an atom that scatters the electron, $f_I(\tilde{q})$ is the atomic form-factor build up by core-shell electrons without the 'free' valence electrons, $Z_I = Z - N_{\text{VB}}$; $N_{\text{VB}}$ is the number of valence electrons per atom or element of a compound; $\rho(\omega,\tilde{q})$ is the electron cloud of the 'free' electrons around the ion core [30]. We assume that both quantities, defining the induced charge *via* the screening of the nucleus by the electrons, respond to some common momentum in the system of the target and the incident electron, $\tilde{q}$. In the original Chihara's paper [30], this momentum is equal to the transferred one $\tilde{q} = q$, since that work considered scattering of incident photons which are moving with the speed of light. In the presented paper, incident electrons are considered which carry charge and travel at different velocities below the speed of light. The nucleus screening depends on the velocity of the incident electron: the faster the electron – the less screened nuclei it "feels", without the contribution of target/atomic electrons having velocities smaller than that of the incident one [31].

Here, we make an *ad hoc* approximation defining this quantity as follows:

$$\tilde{q} = (q + k_i^e)/2, \quad (6)$$

This approximation is chosen to account for the momenta in both fractions of the entire system of charges – the incident electron momentum (wave vector $k_i^e$) and the momentum transferred to the atomic system of the target, $q$. It allows one to obtain a bare nucleus at large $k_i^e$ and standard screening in the region of the Born approximation, as e.g. in Ref. [31].

As was discussed already by van Hove in Ref. [26], the expression of the kind of Eq. (5) contains two types of multipliers: the scattering potential on an individual particle (in Eq.(5), the dynamically screened Coulomb potential), and the dynamic structure factor defining the collective behavior of the atomic ensemble. At high energies of incident electron ($E \gg 100$ eV), the ion-ion dynamic structure factor of the atomic system $S_{ii}(\omega, q)$ reduces to the static structure factor of the ensemble, and the electron scatters on individual non-interacting atoms. This is the typical approximation used in Monte Carlo simulations of transport of fast charged particles [14,32]. However, at low energies ($E \ll 100$ eV), the atomic dynamic structure factor significantly modifies the *e-a* scattering cross section of scattering, and independent atom approximation does not describe the solid target adequately. Low-energy scattering is typically described as the electron-phonon scattering, accounting for collective modes of the atomic dynamics. This approximation is commonly used in the semiconductor physics and optical-laser irradiation approaches [15,33]. Eq.(5) naturally unifies the two limiting cases, containing both terms, the individual atom as well as the collective dynamic structure factor, accounting for the changing dynamical screening [26,27,29].

## 2. Model of atomic loss function

The ion-ion (or atomic) dynamic structure factor $S_{ii}(\omega, q)$, entering Eq.(5), may be calculated with the classical molecular-dynamics simulations [34,35]. Alternatively, the fluctuation-dissipation theorem can be used that in thermal equilibrium links the dynamical structure factor of an ensemble of charged particles $S(\omega, q)$ with its loss function – the imaginary part of the inverse complex dielectric function [36]:

$$Im\left[-\frac{1}{\varepsilon(\omega, q)}\right] = \frac{4\pi e^2}{q^2 \hbar}\left(1 - e^{-\frac{\hbar\omega}{T}}\right) S(\omega, q). \tag{7}$$

For a target consisting of atoms and electrons, Eq. (7) expands to the following:

$$Im\left[-\frac{1}{\varepsilon(\omega, q)}\right] = \frac{4\pi e^2}{q^2 \hbar}\left(1 - e^{-\frac{\hbar\omega}{T}}\right)\left(S_{ee}(\omega, q) - ZS_{ei}(\omega, q) - ZS_{ie}(\omega, q) + Z^2 S_{ii}(\omega, q)\right), \tag{8}$$

Chihara decomposition (Eq.(5)) allows us to obtain the atomic part of the loss function [30]:

$$Im\left[-\frac{1}{\varepsilon_{at}(\omega, q)}\right] = \frac{4\pi e^2}{q^2 \hbar}\left(1 - e^{-\frac{\hbar\omega}{T}}\right)\left(Z - Z_I f_I(\tilde{q}) - \rho(\omega, \tilde{q})\right)^2 S_{ii}(\omega, q). \tag{9}$$

The loss function of the atomic system of a solid can be reconstructed from the experimental optical data on the low-energy photon scattering [37,38] (with $\tilde{q} = q$) whose cross section is proportional to the DSF of the atomic system of a solid [30]:

$$\left(\frac{d^2\sigma}{d(\hbar q)d(\hbar\omega)}\right)^{opt}_{at,exp} \sim Im\left[\frac{-1}{\varepsilon^{opt}_{at,exp}(\omega, q)}\right] = \frac{4\pi e^2}{q^2 \hbar}\left(1 - e^{-\frac{\hbar\omega}{T}}\right) Z^2_{opt} S_{ii}(\omega, q), \tag{10}$$

where the screened charge of atoms in Eq. (10) can be considered constant in the low-energy limit (denoted $Z_{opt}$).

It must be considered that the effective charge of an atom involved in the photon absorption measured in the optical experiments, $Z_{opt}$, differs from that for the scattering of electrons $Z_{e-at} = (Z - f_I(\tilde{q}) - \rho(\tilde{q}))$. Indeed, the target's reaction to low-energy photon irradiation does not include the dependence of atomic charges on the photon momentum (required in Eq. (9)), or information on acoustic phonons. This results in a difference between the experimental atomic loss function obtained from the optical data $Im\left[\frac{-1}{\varepsilon^{opt}_{at,exp}(\omega,q)}\right]$ and the atomic loss function $Im\left[-\frac{1}{\varepsilon_{at}(\omega,q)}\right]$ describing the scattering of an electron on the atomic ensemble of a target. It is similar to the atomic form factors restored from photon, electron or neutron scattering: they are different being restored from the cross sections of different scattering channels (see, e.g., the Mott–Bethe formula connecting the atomic form factors of electron and photon scattering [39,40]).

The charge $Z_{opt}$ in the units of electron charge can be calculated from the f-sum-rule for the experimental optical loss function[23]:

$$Z^2_{opt} = \frac{2}{\pi\hbar^2\Omega_p^2} \int_0^\infty Im\left[\frac{-1}{\varepsilon^{opt}_{at,exp}(\omega,q)}\right](\hbar\omega)d(\hbar\omega), \qquad (11)$$

where $\Omega_p^2 = 4\pi e^2 \sum_i n_i/M_i$, $n_i$ is density of atoms of the i-th type in the compound, and $M_i$ are their masses (for example, in $Al_2O_3$ $Z_{opt} \approx 0.81$, and in $SiO_2$ $Z_{opt} \approx 0.86$).

Comparing Eqs.(9) and (10), we obtain the connection between the calculated loss function for e-a scattering and the one reconstructed from the optical data:

$$Im\left[\frac{-1}{\varepsilon_{at}(\omega,q)}\right] = \frac{(Z - f_I(\tilde{q}) - \rho(\tilde{q}))^2}{Z^2_{opt}} Im\left[\frac{-1}{\varepsilon^{opt}_{at,exp}(\omega,q)}\right]. \qquad (12)$$

Eqs.(11-12) enable rewriting the cross section of scattering of electrons on the atomic ensemble of a target (Eq.(5)) via the loss function reconstructed from the optical data:

$$\frac{d^2\sigma_{e-at}}{d(\hbar q)d(\hbar\omega)} = \frac{2e^2}{n_i\pi\hbar^2 v^2} \frac{1}{\hbar q}\left(1 - e^{\frac{\hbar\omega}{k_B T}}\right)^{-1} \times$$
$$\frac{(Z - f_I(\tilde{q}) - \rho(\omega,\tilde{q}))^2}{Z^2_{opt}} Im\left[\frac{-1}{\varepsilon^{opt}_{at,exp}(\omega,q)}\right]. \qquad (13)$$

Denoting the renormalized loss function as $Im\left[\frac{-1}{\tilde{\varepsilon}_{at}(\omega,q)}\right] = \frac{1}{Z^2_{opt}} Im\left[\frac{-1}{\varepsilon^{opt}_{at,exp}(\omega,q)}\right]$, the e-a scattering cross-section turns into:

$$\frac{d^2\sigma_{e-at}}{d(\hbar q)d(\hbar\omega)} = \frac{2e^2}{n_i\pi\hbar^2 v^2} \frac{1}{\hbar q}\left(1 - e^{\frac{\hbar\omega}{k_B T}}\right)^{-1} (Z - f_I(\tilde{q}) - \rho(\omega,\tilde{q}))^2 Im\left[\frac{-1}{\tilde{\varepsilon}_{at}(\omega,q)}\right]. \qquad (14)$$

The induced charge $\rho(\omega,\tilde{q})$ associated with the electronic cloud of 'free' electrons may be expressed via the valence-band part of the complex dielectric function (CDF)[41]:

$$\rho(\omega, \tilde{q}) = N_{\text{VB}} \left(1 - \frac{1}{|\varepsilon_{VB}(\omega, \tilde{q})|}\right), \tag{15}$$

producing the final expression for the electron elastic scattering in a material:

$$\frac{d^2 \sigma_{e-at}}{d(\hbar q)d(\hbar \omega)} = \frac{2e^2}{n_i \pi \hbar^2 v^2} \frac{1}{\hbar q} \left(1 - e^{\frac{\hbar \omega}{k_B T}}\right)^{-1} \times$$

$$\left[Z - Z_I f_I(\tilde{q}) - N_{\text{VB}} \left(1 - \frac{1}{|\varepsilon_{VB}(\omega, \tilde{q})|}\right)\right]^2 Im\left[\frac{-1}{\tilde{\varepsilon}_{at}(\omega, q)}\right]. \tag{16}$$

The atomic form factors for neutral atoms are available in the literature [12,42], however further approximation may be needed to treat the contributions of ions without the valence electrons in solids. Following the typical shapes of the form-factors of ions [43], we assume the following expression: $Z_I f_I(\tilde{q}) = min(Z_I, Z_I f_a(\tilde{q}))$ (where $f_a(\tilde{q})$ is the atomic form factor of a neutral atom with the same atomic number $Z = Z_I$). Here we use the form factors from Ref.[12].

The used practical model for the valence-band electronic complex dielectric and loss functions will be discussed in the next section.

We note that Eq.(16) recovers known limiting cases. For a scattering of an electron on an isolated atom ($N_{\text{VB}} = 0$, $Z_I = Z$, and assuming $\tilde{q} = q$), the standard scattering cross section is restored [44]:

$$\frac{d^2 \sigma_{e-at}}{d(\hbar q)d(\hbar \omega)} \sim (Z - Z f_a(q))^2.$$

Note that the same limit without dynamical screening effects is reached for fast particles, which interact with a solid as with an ensemble of independent ions[26].

For the case of a fully ionized plasma, ($N_{\text{VB}} = Z$, $Z_I = 0$), the standard electron-ion scattering is recovered (again, assuming $\tilde{q} = q$):

$$\frac{d^2 \sigma_{e-at}}{d(\hbar q)d(\hbar \omega)} \sim \left(\frac{Z}{\varepsilon(\omega, q)}\right)^2.$$

### 3. Model of screening by valence-band electrons

It is, in principle, possible to calculate the valence-band complex dielectric function of a multicomponent system, entering Eq.(16), with *ab initio* methods, such as density-functional based Kubo-Greenwood approach or more advanced techniques [45,46]. However, for practical applications, e.g. in MC simulations, a convenient analytical model of the valence-band complex dielectric function is required. We propose to use the formalism developed by Ritchie and Howie to reconstruct the electronic loss function from the optical coefficients in the form of a sum of

Drude-Lorentz-type oscillators[23]. When $q=0$ (in the optical limit), it can be approximated as follows:

$$Im\left[\frac{-1}{\varepsilon_{VB}(\omega, q=0)}\right] = \sum_{i=1}^{N^{os}} \frac{A_i \gamma_i \hbar \omega}{[\hbar^2 \omega^2 - E_{0i}^2(q=0)]^2 + (\gamma_i \hbar \omega)^2}, \quad (17)$$

here $N^{os}$ oscillators are used to reproduce the valence band part of the loss function from experimental data on the optical complex refraction index (see the detailed algorithm description, e.g., in [22,47]); $E_{0i}$ describes the positions of the maximum of *i*-th oscillator, $\gamma_i$ is the oscillator width (loosely associated with the electronic relaxation time), and $A_i$ is *i*-th oscillator weight, defining contribution of *i*-th oscillator into the total loss function and constrained by the sum rules [22,47]. If the optical data are unavailable for the material of interest, further approximations for the coefficients may be made, allowing to approximately evaluate the loss function [38].

The analytical extension into the finite $q>0$ values is then done *via* the replacement $E_{0i}(q) \to E_{0i}(q=0) + \hbar^2 q^2/(2m)$, where $m$ is the mass of the scattering centre (the electron mass $m_e$ in $\varepsilon_{VB}(\omega, q)$ or the target atom mass in $\varepsilon_{at}(\omega, q)$) [22,48].

Having the loss function, the real part of the inverse CDF may be restored in the analytical form using Kramers-Kronig relations:

$$Re\left[\frac{-1}{\varepsilon_{VB}(\omega, q)}\right] = -\left(1 - \sum_{i=1}^{N^{os}} \frac{A_i(E_{0i}^2(q) - (\hbar \omega)^2)}{[\hbar^2 \omega^2 - E_{0i}^2(q)]^2 + (\gamma_i \hbar \omega)^2}\right), \quad (18)$$

and the complex dielectric function is then recovered:

$$Re[\varepsilon_{VB}(\omega, q)] = -\frac{Re\left[\frac{-1}{\varepsilon_{VB}(\omega, q)}\right]}{Re\left[\frac{-1}{\varepsilon_{VB}(\omega, q)}\right]^2 + Im\left[\frac{-1}{\varepsilon_{VB}(\omega, q)}\right]^2},$$

$$Im[\varepsilon_{VB}(\omega, q)] = \frac{Im\left[\frac{-1}{\varepsilon_{VB}(\omega, q)}\right]}{Re\left[\frac{-1}{\varepsilon_{VB}(\omega, q)}\right]^2 + Im\left[\frac{-1}{\varepsilon_{VB}(\omega, q)}\right]^2}. \quad (19)$$

Note that in the case of a single oscillator $N^{os} = 1$, the CDF reduces to a simpler analytical expression of the Drude-like oscillator:

$$\varepsilon_{VB}(\omega, q) = 1 + \frac{A_1}{E_{01}^2(q) - A_1 - (\hbar \omega)^2 - i\gamma_1 \hbar \omega}. \quad (20)$$

Eqs. (19) or (20) can be used to evaluate the contribution to the screening by the valence-band electrons in Eq.(16).

Summarizing, the *e-a* scattering cross section of an electron on an ensemble of atoms in a solid within the first-order Born approximation with modified dynamical screening is derived, Eq.(16). It contains the atomic and ionic form factors, which may be found in existing databases, e.g. [49,50]. The valence-band electron screening of the target ions may be evaluated *via* the complex dielectric function, restored from optical data or further approximations, reconstructed in the Ritchie-Hovie formalism, using Eqs.(17,19). The loss function of the atomic ensemble may also be reconstructed from the optical data or approximations, Eq.(12).

### III. Results

The above-described formalism was implemented in TREKIS-3 code [51] to evaluate the electron elastic mean free paths (*e-a* scattering) in a variety of materials and compare the results with those from other approaches. The calculated *e-a* scattering mean free paths (MFP) in a few materials are presented below.

We start by discussing a few elemental solids and then move to compounds. For those materials, for which the phonon CDF coefficients are unknown, we used the single-pole approximation of the loss function with the coefficients approximated from the characteristic phonon frequency and the sum rules [38] (such as metals and Si).

For comparison, we will use the approximation proposed for dielectric compounds in [37] (and used in our earlier papers [21,48,52]) where the screened charge in Eq.(14) was replaced by:

$$\left( Z - Z_I f_I(\tilde{q}) - N_{\text{VB}} \left( 1 - \frac{1}{|\varepsilon_{VB}(\omega,\tilde{q})|} \right) \right) \to 1, \qquad (21)$$

and $Im\left[\frac{-1}{\tilde{\varepsilon}_{at}(\omega,q)}\right] \to Im\left[\frac{-1}{\varepsilon_{at,exp}^{opt}(\omega,q)}\right]$. In this case, electrons scatter on atoms with the "effective" charge corresponding to the loss function extracted from scattering of low-energy photons. Below in figures, we mark the results of such calculations as "Z=1". This approximation is expected to work well at low energies, where the electron scattering on atomic ensemble is described by electron-phonon scattering.

Alternatively, we also proposed to describe the effective charge of a target atom by a formula similar to the Barkas one developed for the effective charge of an incident ion in matter [38]:

$$\left( Z - Z_I f_I(\tilde{q}) - N_{\text{VB}} \left( 1 - \frac{1}{|\varepsilon_{VB}(\omega,\tilde{q})|} \right) \right) \to$$
$$\to Z_B = 1 + (Z-1)\left\{ 1 - \exp\left[ -\frac{v}{v_0}(Z-1)^{-\frac{2}{3}} \right] \right\} \qquad (22)$$

where *v* is the incident electron velocity, and $v_0=c/125$ is empirically adjusted atomic electron velocity [53]. Eq.(22) was constructed to reproduce the case Z=1 at low electron energies but reduce the screening to the scattering on a bare nucleus at high electron energies. Calculation with this effective charge will be marked in the plots as "Barkas-like". This expression reduces to Z=1 at the low-energy limit independently of the material. Thus, such an approximation is expected to work reasonably well at both limiting cases, low- and high-energies, but may overestimate the scattering cross section at the intermediate electron energies (see below).

We also compare our results with the often-used screened Rutherford cross section with modified Molier screening parameter (marked in the plots below as "R-M")[14]. This is an example of an independent-atom approximation applicable at high electron energies.

In silicon, shown in Figure 1, the calculated data are compared with the few other calculations from the literature. At low energies, the calculation agree reasonably well with the electron-phonon calculations from [54] and [55]. At high energies, the cross section (and, correspondingly, the mean free path) reproduces the nuclear charge with the reducing screening *via* Barkas-like empirical formula from [38]. At energies above some 10 keV, both reduce to the scattering on unscreened nuclear charge. In this high energy limit, the scattering is reasonably close to the scattering on an isolated atom, which is reproduced with the adjusted R-M, and to the NIST database [56]. The NIST data on the electron elastic scattering on isolated atoms are based on the Dirac partial wave method, which is one of the best approaches available; however, it assumes an independent atom approximation. As mentioned above, in a solid, lattice structure and collective effects dominate for incident electron energies below some ~100 eV, which are not accounted for in independent-atom approaches. Thus, the NIST data are only applicable at high energies (in the case of solids).

Figure 1 demonstrates that the derived Eq.(16) with the used parameters for the phononic and valence band of the loss function reproduces reasonably well both limiting cases, at low energy (electron phonon scattering) and at high energies (electron scattering on an individual ion, close to NIST data and R-M cross section). In contrast, the approximations of Z=1 and Barkas-like expression fail to reproduce the low- and intermediate energy limit; this is the reason why atomic cross sections are often replaced with the electron-phonon scattering cross sections at low energies in MC simulations.

We note that, despite qualitative agreement, there is still some quantitative difference between both CDF-based methods (Eq.(16) and Barkas-like expression) and NIST data at high energies. Since both CDF-based methods produce comparable difference, it indicates that they have a common reason. It may be due to the difference between the dynamic structure factor of the medium in Eq.(16) and the dynamic structure factor of the homogeneous ideal gas in the zero temperature limit [57], which is implied by using the NIST and R-M cross sections (independent atom approximation). It may also be the limitation of the first-order Born approximation, and further development of theory may work on higher-order approximations or methods beyond the perturbation theory to check this possibility.

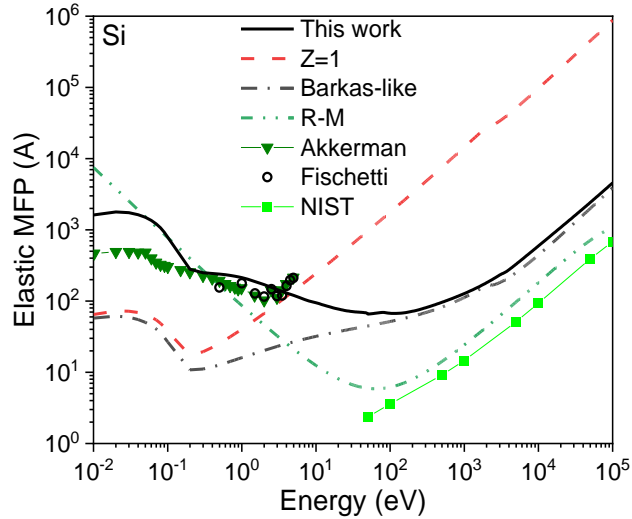

*Figure 1. Mean free path of electron e-a scattering in silicon, calculated with Eq.(16) and compared with other approximations: the calculated electron-phonon scattering from Akkerman and Murat et al.* [54]*, Fischetti and Laux* [55]*, and electron-atom scattering at high energies from NIST database* [56]*.*

Figure 2 shows the calculated *e-a* scattering MFP in aluminum, where again, the high-energy limit is reasonably close to the R-M cross section (independent atom approximation), NIST database [56], and phase-shift calculations by Dapor *et al*. [58] (which used similar method to that of NIST). The small difference again may be due to the difference between the DSF of the medium and the homogeneous ideal gas, as mentioned above. At lower energies, the full-screening expression (Eq.(16)) predicts noticeably larger mean free paths, in contrast to Z=1 and Barkas-like approximations that produce MFPs that strike us as unrealistically short. The MFP calculated with Eq.(16) is closer to those in silicon at low energies, reported above. Thus, it seems, the proposed methodology produces reasonable cross sections over the entire energy range, from low to high electron energies. Unfortunately, there are no experimental data to compare with.

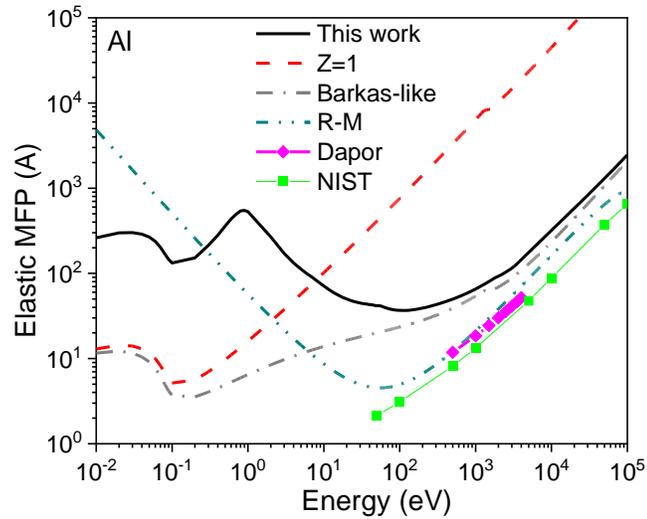

*Figure 2. Mean free path of electron e-a scattering in aluminum, calculated with Eq.(16), compared with other approximations, and the calculated electron-atomic scattering at high energies by Dapor* [58]*, and from NIST database* [56]*.*

Figure 3 shows the calculated *e-a* scattering MFP in gold, where again, the high-energy limit is reasonably close to the R-M cross section, NIST database [56], and phase-shift calculations by Dapor *et al*. [58]. We note that in this case, our calculations agree with NIST database and Dapor's results even better than R-M cross section. The low-energy limit coincides with the Z=1 and Barkas-like expressions in the case of gold.

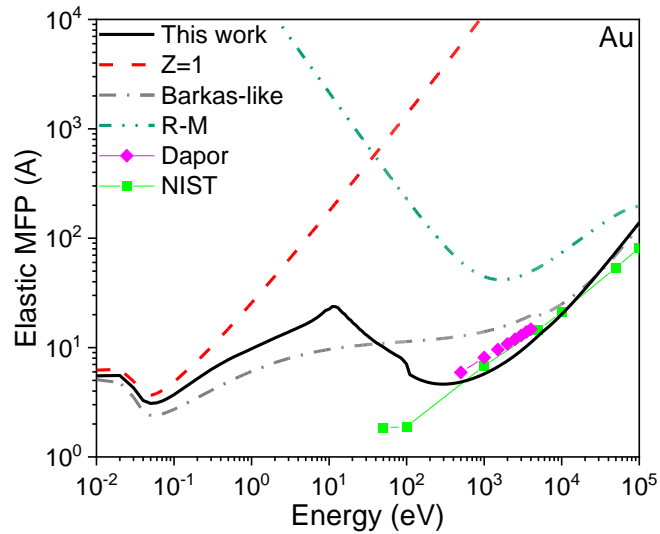

*Figure 3. Mean free path of electron e-a scattering in gold, calculated with Eq.(16), compared with other approximations, and the calculated electron-atomic scattering at high energies by Dapor* [58]*, and from NIST database* [56]*.*

Figure 4 shows the calculated *e-a* scattering MFP in SiC. As in the materials above, the high-energy limit is reasonably close to the R-M cross section (high-energy limit of independent atom approximation). In this case, the R-M cross section was constructed for the compound using the additivity rule summing up the atomic cross sections with the weights defined by the stoichiometry of the compound [59]. In contrast to such atomic calculations, the methodology of Eq.(16) does not rely on the additivity rule, but describes the collective behavior (structure and dynamics) of the solid *via* the dynamic structure factor and the complex dielectric functions. Collective atomic dynamics (and screening by valence electrons of a solid) makes a significant difference at the electron energies below ~100 eV, see Figure 4.

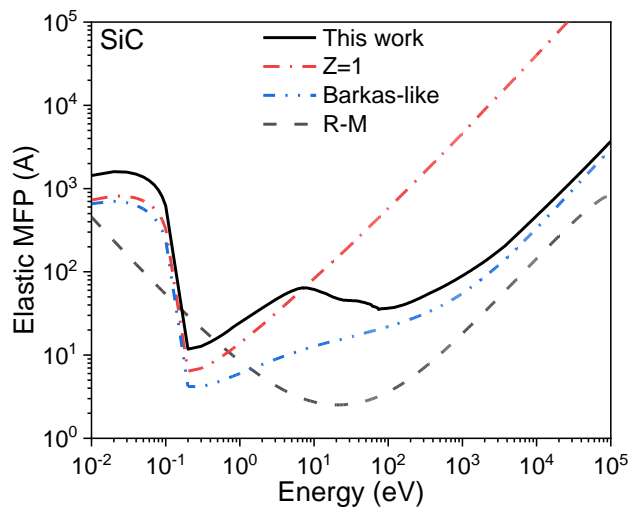

*Figure 4. Mean free path of electron e-a scattering in SiC, calculated with Eq.(16), compared with other approximations.*

Figure 5 shows the calculated *e-a* scattering MFP in $SiO_2$. At low energies, we also compare our results with calculations by Kuhr and Fitting [37], which used CDF-based formalism similar to ours with $Z=1$ approximation – which, thus, resembles this limiting case. Their more recent version of MC modeling used scattering of electrons on optical and acoustic phonons in $SiO_2$ (in Figure 5 added via the Matthiessen rule), which shows an even closer agreement with our calculations [60]. Calculations by Fischetti et al. are also close to those at low energies, but have a deeper drop at electron energies above ~3 eV [61].

At high energies, our calculated MFP coincides with the Barkas-like approximation, and is reasonably close to the R-M cross sections with additivity rule, which almost coincides with Dapor's atomic calculations [62]. Considering the differences among various models in the literature, our agreement with the limiting cases seems reasonable.

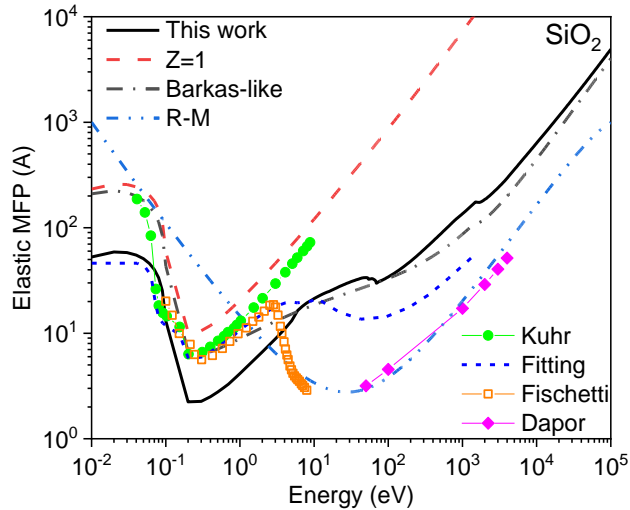

*Figure 5. Mean free path of electron e-a scattering in SiO$_2$, calculated with Eq.(16), compared with other approximations by Kuhr and Fitting [37], Fitting et al. [60], Fischetti et al. [61]; and Dapor's calculations [62].*

Figure 6 shows the calculated *e-a* scattering MFP in Al$_2$O$_3$, where again, the high-energy limit coincides with the Barkas-like approximation and is reasonably close to the R-M cross sections with additivity rule. The remaining difference, again, seems to be either due to the DSF of the medium, or stem from the first-order Born approximation underlying our model. At low energies, the calculated MFPs are close to those from Barkas and Z=1 approximations. The intermediate energies, however, are different among all the cases. We assume that Eq.(16) should produce the best results in this energy range, since it accounts for both, collective atomic dynamics and electronic screening, but so far there is no experimental data to validate our prediction.

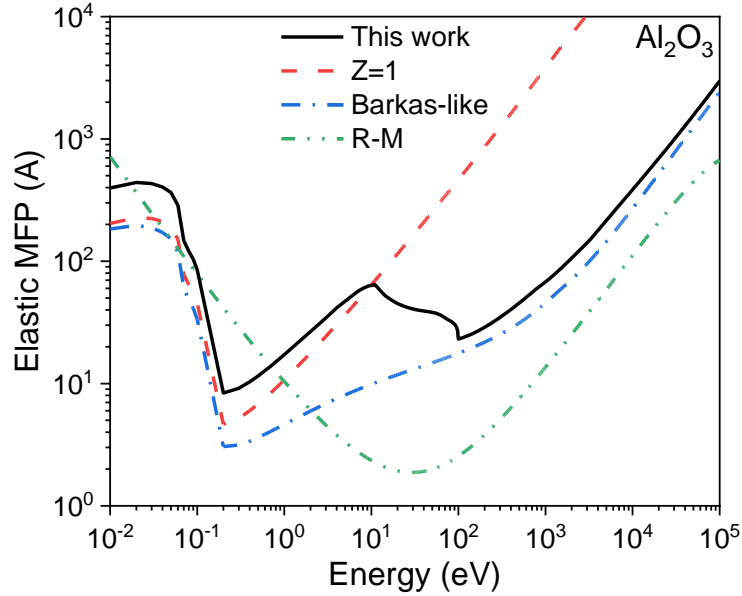

*Figure 6. Mean free path of electron e-a scattering in $Al_2O_3$, calculated with Eq.(16), compared with other approximations.*

Summarizing, the proposed model of *e-a* scattering in matter describes a wide range of energies in a satisfactory agreement with both limiting cases: electron-phonon scattering at low energies and scattering on an individual atom at high energies of the incident electron. Such a unified model takes into account the dependencies of the dynamical screening of the Coulomb interaction and collective response of the atomic ensemble on the velocity of an incident electron. It, thus, absolves the user from seeking different approximations, allowing for a more consistent modeling.

In the next section, we show an example of the model implementation in simulations of electrons and energy transport after swift heavy ion impacts on α-quartz ($SiO_2$) followed by structural changes in the material lattice. The available experimental data, Refs. [63,64], on α-quartz irradiation allow us to validate the developed model.

## IV. Discussion

When a swift heavy ion (SHI; with masses $> 10m_p$ ($m_p$ is the proton mass) and energies 1-10 MeV/nucleon) passes through a matter, it is decelerated in the electronic stopping regime [9]. The resulting electronic excitation relaxes during ~100 fs, which is comparable to the shortest times of atomic ensemble dynamics [9]. During the electron relaxation process, part of the excitation energy is transferred to the lattice. The heated lattice responds within several hundred picoseconds with the formation of a structurally modified region – an SHI track. This separation of scales allows us to model the lattice heating process separately from the lattice relaxation process [9].

We use the transport Monte-Carlo code TREKIS-3 [51] to model the evolution of electron excitation and lattice heating in the vicinity of the ion trajectory. The energy transfer from excited electrons to the lattice in this code may be described with various *e-a* scattering cross sections, including those developed in this work.

The radial profiles energy density transferred to the atomic lattice around the trajectories of 62 MeV (dE/dx=13.4 keV/nm from TREKIS-3) and 1 GeV (dE/dx=24.3 keV/nm) Pb ions obtained with the proposed model (see Eq.(16)) are shown in Figure 7. For comparison, we performed similar calculations with *e-a* cross sections approximations "R-M", "Z=1", and "Barkas-like". As in the previous studies [65,66], the profiles in Figure 7 account for three channels of energy transfer: elastic (*e-a*) scattering of electrons, scattering of valence holes, and nonthermal energy transfer due to electronic excitation. All profiles are averaged over $10^3$ MC iterations for reliable statistics. It can be seen that the calculation results strongly depend on the choice of the *e-a* scattering model.

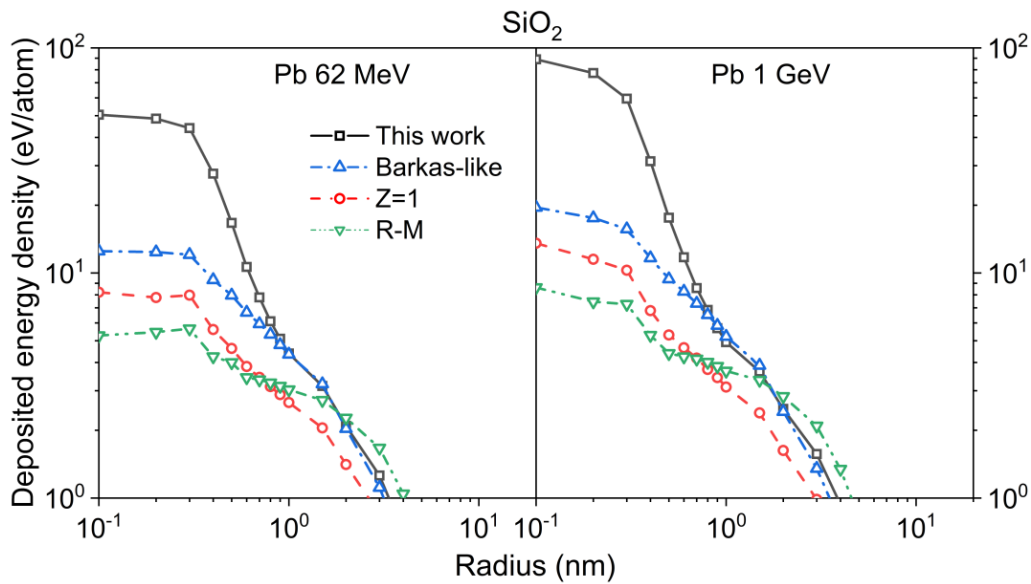

*Figure 7. Density of the energy transferred to atoms around the trajectory of 62 MeV (left panel) and 1 GeV (right panel) Pb ions in $SiO_2$, evaluated with different approximations for e-a scattering cross sections.*

To predict structural changes in the tracks of Pb ions with energies of 62 MeV and 1 GeV in $SiO_2$, the profiles of the energy deposited in the lattice (see Figure 7) were converted into initial atomic velocities using the algorithm proposed in Ref. [52]. We calculated atomic velocities in cylindrical layers around the ion trajectory assuming Gaussian-like dispersion of kinetic energy and uniform distribution of atomic momenta within the solid angle. Further trajectories of atoms were modeled by classical molecular dynamics method in the LAMMPS simulation package [67] using the Vashishta interatomic potential [68]. In the simulations, we used a supercell contains 533,520 atoms with a size of 25x25x10 nm$^3$ with periodic boundary conditions. Throughout the simulation, the Berendsen thermostat at room temperature, with a characteristic damping time of 100 fs, was applied to the box walls around the SHI trajectory in a 0.5 nm thick layer. The simulations were performed until the temperature of the irradiated region in the supercell dropped below 500 K after the initial excitation, which typically required ~100 ps for the studied cases.

Results of the simulation of the atomic response to such energy deposition profile (see Figure 7) obtained with different *e-a* cross sections are shown in Figure 8. The observed tracks are amorphous, which is in agreement with experimental data [63,64]. The experimental value of the 62 MeV Pb track diameter in $SiO_2$ is 11±2 nm [63]. It can be seen that the diameter of 8.5 nm closest to the experimental value corresponds to the profile calculated with the cross section Eq.(16) proposed in this work. The other calculations underestimate the size of the amorphous region: in "R-M", this is due to the exclusion of scattering on phonons; in the case of "Z=1", the exclusion of the effect of reducing of the screening of atomic nuclei at high energies overestimates the MFP; and in the "Barkas-like" case, the neglect of the difference between the screening by valence electrons and the screening by shell electrons (only shell electrons are considered in the model).

For irradiation of $SiO_2$ with 1 GeV Pb ion (at the Bragg peak of the ion electronic energy loss), the experimental track diameter is 9.6±1 nm [64]. It can be seen that the profile from Figure 7 obtained with the cross section proposed in this work (Eq.(16)) produces the value of ~9.5 nm in an excellent agreement with the experiment.

The fraction of high-energy particles in the spectrum of excited electrons increases with the ion energy [69]. For such short-wavelength electrons, the approximation of instantaneous scattering on independent lattice atoms works well. Therefore, the profile obtained using the R-M cross section also produces a close to the experimental value of the track diameter of ~8.9 nm.

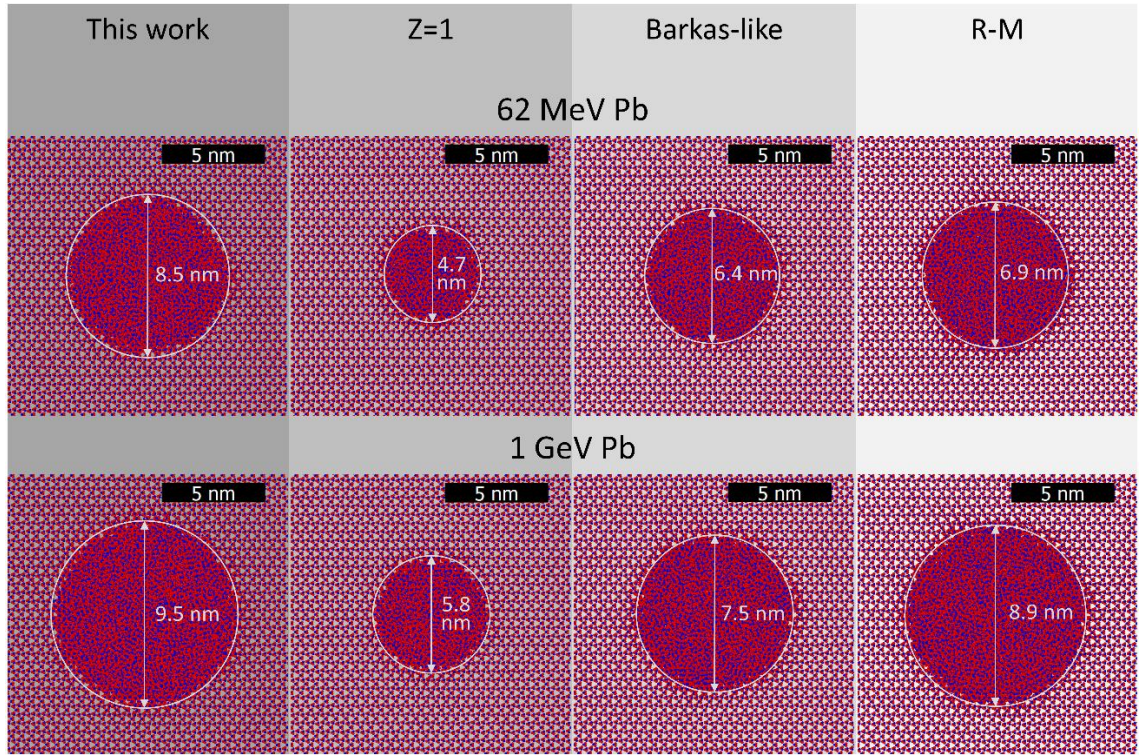

*Figure 8. Snapshots of formed tracks in SiO$_2$ irradiated with 62 MeV Pb ion (top panel) and 1 GeV Pb ion (bottom panel) obtained with different approximations for e-a scattering cross sections.*

The fact that the track sizes for Pb at energies of 62 MeV and 1 GeV calculated using Eq. (16) agree well with the experimental data – better than other models used – is a strong argument in support of the developed model of the *e-a* electron scattering cross section.

## V.     Conclusions

A new model of the cross section of electron non-ionizing scattering on the atomic system (*e-a* scattering) in condensed matter is proposed. It is based on the dynamic structure factor (linear response theory) formalism. Accounting for the dynamical screening of target atoms, the developed cross section unifies the limiting cases of the electron-phonon scattering in the low-energy limit, and electron-ion scattering similar to scattering on the atomic gas in the high-energy one. Comparison with the available simulations shows a reasonable agreement in both limits in various materials. The formalism enables a straightforward implementation, e.g., in Monte Carlo simulations. The profiles of energy deposited in the SiO$_2$ lattice by electrons excited by the Pb ion with energies of 62.8 MeV and 1 GeV calculated with this formalism agree well with the available experimental data [63,64] validating the application of the proposed model.

## VI. Acknowledgements

NM thanks the financial support from the Czech Ministry of Education, Youth, and Sports (grant nr. LM2023068), and the computational resources provided by the e-INFRA CZ project (ID:90254), supported by the Ministry of Education, Youth and Sports of the Czech Republic. DIZ's work was carried out using computing resources of the federal collective usage center Complex for Simulation and Data Processing for Mega-science Facilities at NRC "Kurchatov Institute", http://ckp.nrcki.ru/.

## VII. Code and data availability

The code TREKIS-3 used to obtain the results presented in this study is available from [51].

http://dx.doi.org/10.48550/arXiv.2405.06950

dynamics analysis, Comput. Phys. Commun. 279 (2022) 108454. https://doi.org/10.1016/j.cpc.2022.108454.

[35] S.A. Gorbunov, P.N. Terekhin, N.A. Medvedev, A.E. Volkov, Combined model of the material excitation and relaxation in swift heavy ion tracks, Nucl. Instruments Methods Phys. Res. Sect. B Beam Interact. with Mater. Atoms 315 (2013) 173–178. https://doi.org/10.1016/j.nimb.2013.04.082.

[36] C. Kittel, Quantum theory of solids, 1963.

[37] J.-C. Kuhr, H.-J. Fitting, Monte Carlo simulation of electron emission from solids, J. Electron Spectros. Relat. Phenomena 105 (1999) 257–273. https://doi.org/10.1016/S0368-2048(99)00082-1.

[38] N. Medvedev, F. Akhmetov, R.A. Rymzhanov, R. Voronkov, A.E. Volkov, Modeling time-resolved kinetics in solids induced by extreme electronic excitation, Adv. Theory Simulations 5 (2022) 2200091. https://doi.org/10.1002/ADTS.202200091.

[39] The scattering of electrons by atoms, Proc. R. Soc. London. Ser. A, Contain. Pap. a Math. Phys. Character 127 (1930) 658–665. https://doi.org/10.1098/RSPA.1930.0082.

[40] H. Bethe, Zur Theorie des Durchgangs schneller Korpuskularstrahlen durch Materie, Ann. Phys. 397 (1930) 325–400. https://doi.org/10.1002/ANDP.19303970303.

[41] N.H. March, M.P. Tosi, Atomic Dynamics in Liquids, Courier Corporation, Chelmsford, 1991. https://books.google.com/books?id=B7gZWY5m6fgC&pgis=1 (accessed April 10, 2015).

[42] D.E. Cullen, EPICS2017: Electron Photon Interaction Cross Sections: w-nds.iaea.org/epics/, Vienna, 2018. https://www-nds.iaea.org/publications/iaea-nds/iaea-nds-224%7B%5C_%7DRev1%7B%5C_%7D2018.pdf.

[43] S.P. Hau-Riege, X-ray atomic scattering factors of low-Z ions with a core hole, Phys. Rev. A 76 (2007) 042511. https://doi.org/10.1103/PhysRevA.76.042511.

[44] L.D. Landau, L.M. Lifshitz, Quantum Mechanics, Third Edition: Non-Relativistic Theory, 3 edition, Butterworth-Heinemann;, 1976. http://www.amazon.com/Quantum-Mechanics-Third-Edition-Non-Relativistic/dp/0750635398 (accessed June 30, 2014).

[45] B. Holst, V. Recoules, S. Mazevet, M. Torrent, A. Ng, Z. Chen, S.E. Kirkwood, V. Sametoglu, M. Reid, Y.Y. Tsui, Ab initio model of optical properties of two-temperature warm dense matter, Phys. Rev. B - Condens. Matter Mater. Phys. 90 (2014) 35121. https://doi.org/10.1103/PhysRevB.90.035121.

[46] L. Calderín, V.V. Karasiev, S.B. Trickey, Kubo–Greenwood electrical conductivity formulation and implementation for projector augmented wave datasets, Comput. Phys. Commun. 221 (2017) 118–142. https://doi.org/10.1016/j.cpc.2017.08.008.

[47] N. Medvedev, Modeling ultrafast electronic processes in solids excited by femtosecond VUV-XUV laser Pulse, AIP Conf. Proc. 582 (2012) 582–592. https://doi.org/10.1063/1.4739911.

[48] R.A. Rymzhanov, N.A. Medvedev, A.E. Volkov, Effects of model approximations for electron, hole, and photon transport in swift heavy ion tracks, Nucl. Instruments Methods